\RequirePackage{lineno}
\documentclass[sort&compress,10pt,5p]{elsarticle}
\usepackage{hyperref} 
\usepackage{atlasphysics} 
\usepackage{preprintcover}

\usepackage{subfigure}
\usepackage{mathrsfs}
\usepackage{graphicx}
\usepackage{multirow}
\usepackage{rotating}
\usepackage{xspace}
\usepackage{slashed}
\usepackage{amsmath}

\journal{Physics Letters B}

\PreprintCoverPaperTitle{Search for TeV-scale Gravity Signatures in Final States with Leptons and Jets with the ATLAS Detector at $\sqrt{s}$ = 7 TeV}  
\PreprintIdNumber{CERN-PH-EP-2012-046}  
\PreprintCoverAbstract{
The production of events with multiple high transverse momentum particles including charged leptons and jets is measured, using \lumi~fb$^{-1}$ of proton-proton collision data recorded by the ATLAS detector during the first half of 2011 at $\sqrt{s} = 7$~TeV. No excess beyond Standard Model expectations is observed, and upper limits on the fiducial cross sections for non-Standard Model production of these final states are set. Using models for string ball and black hole production and decay, exclusion contours are determined as a function of mass threshold and the fundamental gravity scale.
}  
\PreprintJournalName{Physics Letters B}  

\newcommand{\lumi}{1.04} 
\newcommand{\alpgen}{{\sc Alpgen}\xspace}
\newcommand{\herwig}{{\sc Herwig}\xspace}
\newcommand{\mcatnlo}{{\sc MC@NLO}}

\newcommand{\jimmy}{{\sc Jimmy}\xspace}
\newcommand{\pythia}{{\sc Pythia}\xspace}
\newcommand{\sherpa}{{\sc Sherpa}\xspace}
\newcommand{\blackmax}{{\sc Blackmax}\xspace}
\newcommand{\charybdis}{{\sc Charybdis}\xspace}
\newcommand{\geant}{{\sc GEANT4}\xspace}
\newcommand{\sumpt}{\sum p_\mathrm{T}}
\newcommand{\MD}{\ensuremath{M_{\mathrm{D}}}}
\newcommand{\MTH}{\ensuremath{M_{\mathrm{TH}}}}
\newcommand{\MS}{\ensuremath{M_\mathrm{S}}}

\begin{document}
\begin{frontmatter}

\title{Search for TeV-scale gravity signatures in final states with leptons and jets \\ with the ATLAS detector at $\sqrt{s}$ = 7 TeV}

\author{The ATLAS Collaboration}

\begin{abstract}

The production of events with multiple high transverse momentum particles including charged leptons and jets is measured, using 
\lumi~fb$^{-1}$ of proton-proton collision data recorded by the ATLAS detector during the first half of 2011 at $\sqrt{s} = 7$~TeV. 
No excess beyond Standard Model expectations is observed, and upper limits on the fiducial cross sections for non-Standard Model production of these final states are set.
Using models for string ball and black hole production and decay, exclusion contours are determined as a function of 
mass threshold and the fundamental gravity scale.

\end{abstract}
\end{frontmatter}

\section{Introduction}
\label{sec:introduction}

Models proposing extra spatial dimensions address the mass hierarchy problem,
the origin of the sixteen orders of magnitude separation between the electroweak
and Planck scales. These allow the gravitational field to propagate into the $(n+4)$ 
dimensions, where $n$ is the number of extra spatial dimensions, while Standard Model (SM) fields are constrained to lie in our 
four-dimensional brane. Consequently, the 
resulting Planck scale in $(n+4)$ dimensions, $\MD$, is greatly diminished compared to the four-dimensional analogue, $M_\mathrm{Pl}$, and should
be near the other fundamental scale, the electroweak scale, if the hierarchy problem is
to be addressed.
Such low-scale gravity models allow the existence of 
gravitational states such as black holes and, within the context of weakly-coupled
string theory, string balls, that could be produced with appreciable cross sections at the Large Hadron Collider (LHC).

Two such extra-dimensional scenarios are the Randall-Sundrum models~\cite{Randall:1999ee} and the large, flat extra-dimensional ADD models~\cite{ArkaniHamed:1998rs,Antoniadis:1998ig}.
In the large extra dimension
scenario, there are a number $n>1$ of additional flat extra dimensions, and $\MD$ is determined by the volume and
shape of the extra dimensions.  Within the context of this model, experimental lower limits on the value
of $\MD$ have been obtained from experiments at LEP~\cite{LEPADD} and 
the Tevatron~\cite{Aaltonen:2008hh,Abazov:2008kp}, as well as at
ATLAS~\cite{Aad:2011xw,Aad:2011bw} and
CMS~\cite{Chatrchyan:2011nd},
by searching for production of
the heavy Kaluza-Klein gravitons associated with the extra dimensions.  The most stringent limits~\cite{Aad:2011xw}
come from the LHC analyses that search for non-interacting gravitons recoiling against a single jet
(monojet and large missing transverse energy), and range from
$\MD > 2.0$~TeV, for $n=6$, to $\MD > 3.2$~TeV, for $n=2$.  Due to the greatly
enhanced strength of gravitational interactions at short distances, or high energies, the formation
of 
gravitational states such as black holes or string balls at the LHC is another signature of extra dimensional models.

Large extra dimensions can be embedded into weakly-coupled string theory~\cite{Dimopoulos:2001qe,Gingrich:2008di}.
In these models, black holes end their Hawking evaporation phase when their
mass reaches a critical value $\MS/g^2_\mathrm{S}$, also known as the correspondence
point, where $\MS$ and 
$g_\mathrm{S}$ are the string scale and coupling constant, respectively. 
At this point they transform into high-entropy string states -- string
balls -- which, in turn, continue to decay thermally.  

The semi-classical approximations used in the modelling of black hole production are valid only for partonic centre-of-mass energies well above $\MD$, motivating the use of a minimal threshold $\MTH$ to remove contributions where the modelling is not reliable.
The resulting black hole mass distribution ranges from this threshold up to $\sqrt{s}$.
The precise mass value above which the production of such high multiplicity states is feasible is uncertain.
A conservative interpretation~\cite{Anchordoqui01,Gingrich06a}
is that $\MTH > 3\MS$ for string balls and $\MTH > 5\MD$ for black holes.

Thermal radiation is thought to be emitted by black holes due to quantum 
effects~\cite{Hawking:1974sw}. 
A black hole, in $(n+4)$ dimensions, of given mass and angular momentum is characterised by a Hawking temperature, which is higher for a lighter, or more strongly rotating, black hole. 
Grey-body factors modify the spectrum of emitted particles from that of a perfect thermal black body~\cite{Page76a}, by quantifying the transmission probability through the curved space-time outside the horizon; these emissivities depend upon particle spin, $n$ and the properties of the black hole. 
All Standard Model particles are emitted.

As the black hole mass approaches the Planck scale and few further emissions are expected, quantum effects become important and classical evaporation is no longer a suitable description. The remaining black hole remnant is decayed to a small number of SM particles\footnote{This paper considers both high multiplicity, generated by the \blackmax burst model or the \charybdis variable multiplicity decay with a four-body average, and low multiplicity decays to two bodies.}.

Were black hole states\footnote{In this
Letter, all references to black holes also apply to string
balls, unless otherwise stated.} to be produced at the LHC, they would decay to final states with a 
relatively high multiplicity of high-$p_\mathrm{T}$ 
particles, most commonly jets. While the multiplicity is generally high, the exact spectrum is rather model dependent: for example, the inclusion of black 
hole rotation leads to a somewhat lower multiplicity of higher energy emissions~\cite{Frost:2009cf}. 
One of the few more robust predictions of these models is the expectation that particles are produced approximately according to 
their degrees of 
freedom,
(with some modification by the relative emissivities).
This is the ``democratic'' or ``universal'' coupling of gravity. 
Thus, the probability for the production of a leptonic final state varies primarily with the emission multiplicity, which depends upon model parameters and the remnant state treatment. 
Nonetheless, this multiplicity dependence is much reduced compared to using the multiplicity directly, for even low multiplicity decays will frequently contain a lepton.
Hence, these models predict the existence of at least one high-$p_\mathrm{T}$ lepton\footnote{Throughout this Letter ``lepton'' denotes 
electrons and muons only.}
in a significant fraction ($\sim 15-50~\%$) of final states for black holes or string balls with $\MD$ and $\MTH$ values
in the range accessible to LHC experiments and not already excluded.
The largest theoretical uncertainties in the modelling of these states are the limited knowledge of gravitational radiation and the resultant cross section during the formation phase, and the uncertainties of the decay process as the black hole mass approaches $M_D$, especially the treatment of the remnant state.

Searches for these gravitational states 
have previously been performed by investigating final states with multiple
high-$\pt$ objects~\cite{Khachatryan:2010wx},
high-$\pt$ jets only, and in dimuon events~\cite{Aad:2011bw}.
This analysis searches for an excess of multi-object events produced at high
$\sumpt$, defined as the scalar sum of $\pt$ of the reconstructed objects selected (hadronic jets, electrons and muons).
Only events containing at least one isolated electron or muon are selected.
While jets should dominate the decays of black holes, the rate for lepton production is
anticipated to be sizable, as noted above.
Additionally, the requirement of a high-$p_\mathrm{T}$ lepton
significantly reduces the dominant multi-jet background, 
whilst maintaining a high efficiency for black hole events.

This search considers final states with three or more selected objects (leptons or jets), and consequently is not sensitive to two-body final states, as predicted in so-called quantum black hole states~\cite{Meade:2007sz}. Results for two-body final states can be found in Ref.~\cite{Aad:2011aj}.

\section{The ATLAS Detector}

The ATLAS detector~\cite{Aad:2008zzm} is a multipurpose particle
physics apparatus with a forward-backward symmetric cylindrical
geometry and nearly 4$\pi$ coverage in solid angle\footnote{
ATLAS uses a right-handed coordinate system with its origin at the nominal
interaction point in the centre of the detector and the $z$-axis along the
beam pipe. Cylindrical coordinates $(r,\phi)$ are used in the transverse
plane, $\phi$ being the azimuthal angle around the beam pipe. The pseudorapidity $\eta$ is
defined in terms of the polar angle $\theta$ by $\eta=-\ln\tan(\theta/2)$.
}.  
The layout of the detector is
dominated by four superconducting magnet systems, which comprise a
thin solenoid surrounding inner tracking detectors and three large
toroids, each consisting of eight coils. 
The inner detector consists of a silicon pixel detector, a silicon microstrip detector (SCT) and a transition radiation tracker (TRT).
In the pseudorapidity
region $\left|\eta\right| < 3.2$, high-granularity liquid-argon (LAr)
electromagnetic (EM) sampling calorimeters are used.  An iron-scintillator tile
calorimeter provides coverage for hadronic showers over
$\left|\eta\right| < 1.7$.  The end-cap and forward regions, spanning
$1.5~<~\left|\eta\right|~<~4.9$, are instrumented with LAr calorimetry
for both EM and hadronic measurements.
The muon spectrometer surrounds these, and comprises a system of precision tracking chambers, and detectors for triggering.

\section{Trigger and Data Selection}
\label{sec:trigger}

The data used in this analysis were recorded between March and July
in 2011, with the LHC operating at a centre-of-mass energy of $7$~\TeV.
The integrated luminosity is \lumi~fb$^{-1}$, with an uncertainty of
$3.7\%$~\cite{ATLAS-CONF-2011-116,Aad:2011dr}. 

Events are required to pass either a single electron or a single muon trigger, for the electron and muon channels, respectively.  
The electron (muon) trigger threshold lies at transverse energy, $E_\mathrm{T} = 20$~GeV ($p_\mathrm{T} = 18$~GeV).
The trigger efficiencies reach the plateau region for 
lepton transverse momenta values substantially below the minimum analysis threshold of 40~GeV, 
with typical trigger efficiencies for leptons selected for offline analysis of: $96\%$ for electrons~\cite{Aad:2011mk}, 
$75\%$ for muons with $|\eta|<1.05$ and $88\%$ for 
muons with $1.05~<~|\eta|~<~2.0$~\cite{Almond1274252}. 

\section{Monte Carlo Simulation}
\label{sec:mcsamples}

Monte Carlo (MC) simulated event samples are used to develop and validate the analysis procedure, to help estimate the SM backgrounds in the signal region 
and to investigate specific signal models. Jets produced via QCD processes are generated with \pythia~\cite{Sjostrand:2006za}, using the {\sc MRST2007LO*} modified 
leading-order parton distribution functions (PDF) \cite{Sherstnev:2007nd}, which are used with all leading-order (LO) Monte Carlo generators.
The production of top quark pairs and of single top quarks is simulated with \mcatnlo~\cite{mcatnlo} (with a top quark mass of $172.5$ \GeV) and the next-to-leading order (NLO) 
PDF set {\sc CTEQ6.6}~\cite{Nadolsky:2008zw}, which is used with all NLO MC generators. Samples of $W$ and $Z/\gamma^*$ Monte Carlo events 
with accompanying jets are 
produced with \alpgen~\cite{Mangano:2002ea}, using the {\sc CTEQ6L1} PDFs~\cite{Pumplin:2002vw}, and events generated with \sherpa~\cite{Gleisberg:2003xi} are used to assess the systematic uncertainty associated with the 
choice of MC generator.
Diboson ($WW$, $WZ$, $ZZ$) production is simulated
with \herwig~\cite{herwig}. 
Fragmentation and hadronisation for the \alpgen~and \mcatnlo~samples are performed with \herwig, using
\jimmy~\cite{Butterworth:1996zw} for the underlying event. 
All MC samples are produced using a specific ATLAS parameter 
tune~\cite{ATL-PHYS-PUB-2011-008} and the ATLAS full \geant~\cite{geant4} detector simulation~\cite{simulation}. 
The MC samples are produced with a simulation of multiple interactions per LHC bunch crossing (pile-up). 
Different pile-up conditions as a function of the LHC instantaneous luminosity are taken into account by reweighting MC events according to the number of interactions observed in the data, which has a mean of about six.

Signal samples are generated with the \charybdis~\cite{Frost:2009cf} and \blackmax~\cite{Dai:2007ki,Dai:2009by} generators. 
The shower evolution and hadronisation uses \pythia, 
with the {\sc CTEQ6.6} PDF sets using the black hole mass as the QCD scale.
No radiation losses in the formation phase are modelled.
The \charybdis samples are generated with both low and high multiplicity remnants,
whilst the \blackmax samples use the final burst remnant model, which gives high multiplicity remnant states~\cite{Dai:2007ki}.
The high multiplicity options of both generators 
produce concordant distributions.
Samples are generated for both rotating and non-rotating black holes for six extra dimensions.  
Focus is placed on models with six extra dimensions due to the less stringent limits on $\MD$.
String ball samples are produced with \charybdis for both rotating and non-rotating cases, for six extra dimensions, and a string coupling, 
$g_\mathrm{S}$, of $0.4$.
For each benchmark model, samples are generated with $\MD$ ($\MS$ for string ball models) varying from $0.5$--$2.5$~TeV and $\MTH$ from $3$--$5$~TeV.

\section{Object Reconstruction}

\label{sec:objectreconstruction}

Electrons are reconstructed from clusters in the electromagnetic calorimeter
matched to a track in the inner detector~\cite{Aad:2011mk}.  A set of electron
identification criteria based on the calorimeter shower shape, track quality and track
matching with the calorimeter cluster are described in Ref.~\cite{Aad:2010yt} and are referred to
as ``loose'', ``medium'' and ``tight''.
Electrons 
are required to have $\pt > 40 \GeV$, $|\eta|~<~2.47$ and to pass the ``medium'' electron definition. 
Electron candidates are required to be isolated: the sum of the transverse energy deposited within a cone of size $\Delta R < 0.2$ around the electron candidate
(corrected for transverse shower leakage and pile-up from additional {\it pp} collisions) is required to be less than
$10\%$ of the electron $\pt$.
Electrons with a distance to the
closest jet of $0.2 < \Delta R < 0.4$ are discarded, where
$\Delta R~=~\sqrt{(\Delta \eta)^2 + (\Delta \phi)^2}$.

Muon candidates are selected from a combined track in the muon spectrometer and in the inner detector.
Muons are required to have $\pt > 40 \GeV$. Muon candidates are required to have an associated inner detector track with sufficient 
hits in the pixel, SCT and TRT detectors to ensure a good measurement. 
Additional requirements are made on the muon system hits in order to guarantee the best possible resolution at high $\pt$:
muon candidates must have hits in at least three precision layers and no hits in 
detector regions with more limited alignment precision.
These requirements effectively restrict the muon acceptance to the barrel region ($|\eta|<1.0$) and a portion of the end-cap
region ($1.3 < |\eta| < 2.0$)~\cite{Collaboration:2011yg}.
Muons with a distance to the closest jet of $\Delta R < 0.4$ are discarded.
In order to reject muons resulting from cosmic rays, 
requirements are placed on the distance of each muon track from a reconstructed
primary vertex (PV): 
$|z_0| < 1$~mm 
and $|d_0| < 0.2$ mm, where $z_0$ and $d_0$ are the impact parameters of each muon
in the longitudinal and transverse planes, respectively. 
Muons must be isolated: the $\pt$ sum
of tracks within a cone of $\Delta R < 0.3$ around the muon candidate
is required to be less than $5\%$ of the muon $\pt$.

Jets are reconstructed using the anti-$k_t$ jet clustering 
algorithm~\cite{Cacciari:2008gp} with a distance parameter $R$ of $0.4$. 
The inputs to the jet algorithm are clusters seeded
from calorimeter cells with energy deposits significantly above the measured noise~\cite{Lampl:2008}.
Jets are corrected for effects from calorimeter non-compensation and inhomogeneities through the use of $\pt$- and $\eta$-dependent 
calibration factors based on Monte Carlo corrections validated with test-beam and collision data~\cite{jes}.
This calibration corresponds to the scale that would be obtained applying the jet algorithm to stable particles at the primary collision vertex.
Selected jets are required to have $\pt > 40 \GeV$
and $|\eta| < 2.8$. Events with jets failing jet quality criteria
against noise and non-collision backgrounds  
are rejected~\cite{Aad:2011he}. Jets within
a distance $\Delta R < 0.2$ of a selected electron are also rejected.

\begin{figure*}[htb]
\begin{center}
\subfigure[]{
\includegraphics[width=0.40\textwidth]{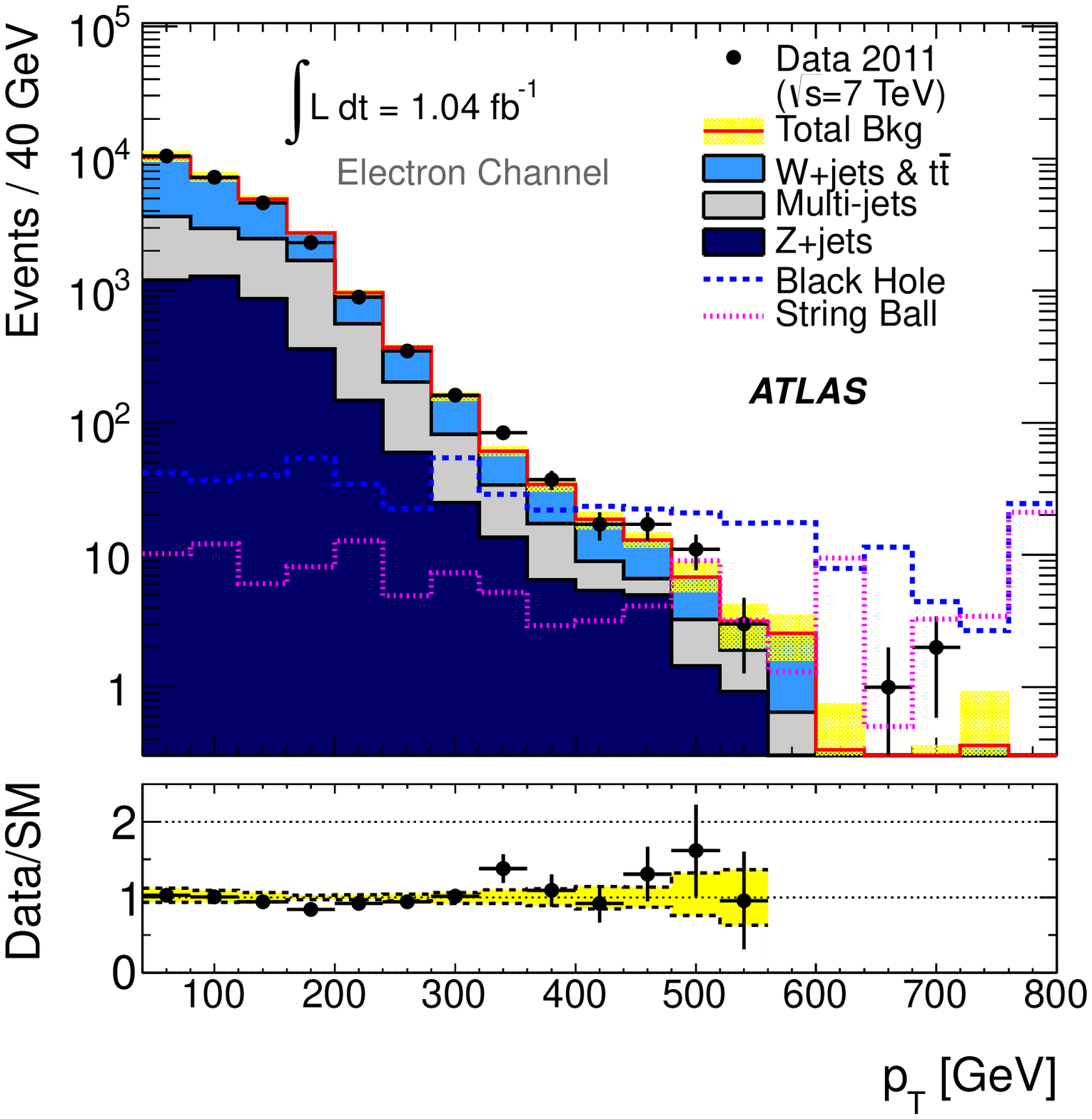}}
\subfigure[]{
\includegraphics[width=0.40\textwidth]{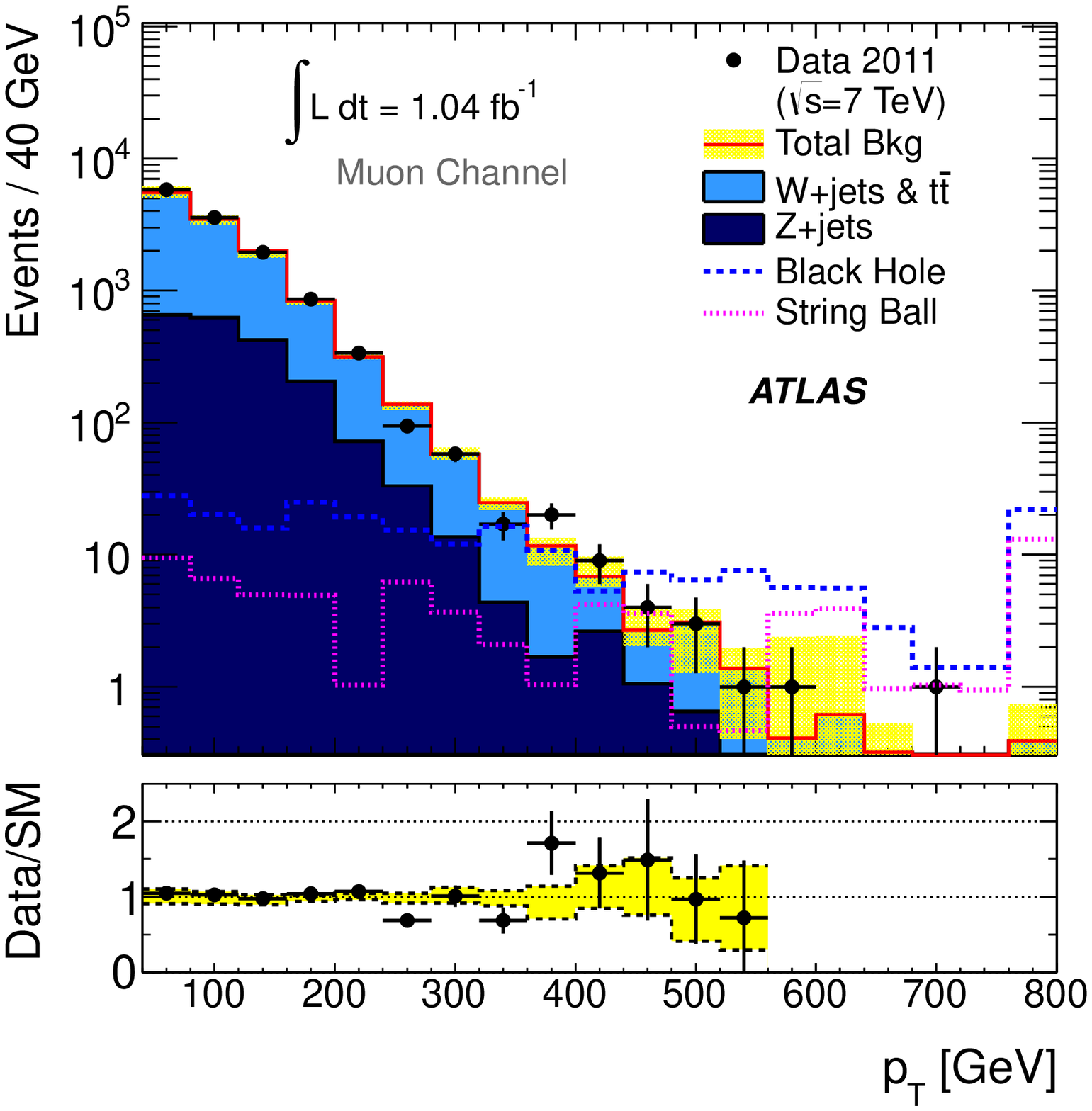}}
\caption{The transverse momentum distributions of the highest momentum lepton, after event preselection, in electron (a) and muon (b) channels. 
The Monte Carlo 
distributions are rescaled to be in agreement with data in selected control regions. 
The lower panels show the ratio of the data to the expected background (points) and the uncertainty (shaded band).
The shaded band in each panel indicates the total uncertainty on the expectation from the finite size of event samples, jet and lepton energy scales and resolutions.
Two representative signal
distributions are overlaid for comparison purposes.  The signal labelled ``Black Hole'' is a non-rotating black hole sample with $n=6$,
$\MD=0.8$~TeV and $\MTH=4$~TeV.  The signal labelled ``String Ball'' is a rotating string ball sample with $n=6$, $\MD=1.26$~TeV,
$\MS=1$~TeV and $\MTH=3$~TeV.  Both signal samples were generated with the \charybdis{} generator.  The final histogram bin shows
the integral of all events with $\pt\ge 760$~GeV. }
\label{fig:prelep} 
\end{center}
\end{figure*}

\section{Event Selection}
\label{sec:eventselection}

Events are required to have a reconstructed primary vertex associated with at least five tracks.
During the data-taking period considered, a readout failure in the LAr barrel calorimeter resulted in a small ``dead'' region, 
in which up to $30\%$ of the incident jet energy may be lost. Should any of the four leading jets with $\pt>40$~GeV fall into this region, 
the event is vetoed. This is applied consistently to all data and Monte Carlo events, and 
results in a loss of signal efficiency of $\sim 15-20\%$ for the models considered.
Additionally, electrons incident on this region are discarded.
Selected events contain at least one high-$\pt$ ($>40$~GeV) isolated lepton. 
Two statistically independent samples are defined by separating events for which the leading lepton (that of highest $\pt$) is 
an electron (muon) into an electron (muon) channel sample. 

High multiplicity final states of interest can be separated from Standard Model background events using the quantity:

\begin{equation}
\sumpt = \sum_{i=\mathrm{objects}} p_{\mathrm{T},i}\, ,
\end{equation}
which is the scalar sum of the transverse momenta of the selected final state reconstructed objects (leptons and jets), described in Section~\ref{sec:objectreconstruction}. 
The signal, containing multiple high-$p_\mathrm{T}$ leptons and jets, manifests itself at high $\sumpt$. 

The missing transverse momentum $\met$ is defined as the opposite
of the vectorial $\pt$ sum of reconstructed objects in the event,
comprising selected leptons, jets with $\pt > 20 \GeV$, 
any additional identified non-isolated muons, and calorimeter clusters not belonging to any of the
aforementioned object types.
Although $\met$ is not considered as an object in this analysis, 
it is used in the definitions of regions for background estimation.

Preselection requirements are used to select a sample of events with similar kinematics and composition to the signal regions for this search, described later in this
Section, but with lower $\pt$ thresholds for selected objects.
Events are required to have at least three selected objects passing the $40$~GeV $\pt$ threshold, with at least one lepton, and have a moderate requirement of $\sumpt$ $> 300$~GeV. 
Additionally, the electron channel requires the leading electron to pass the ``tight'' selection. 
Figure~\ref{fig:prelep} shows the transverse momentum of the leading lepton after event preselection for electron and muon channels,
where the background distributions have been normalised to be in agreement with data in selected control regions, as described in
Section~\ref{sec:backgrounds}.  

For the signal region, the $\sumpt$, lepton and jet $\pt$ requirements are raised further.
Events are required to contain at least three reconstructed objects with $\pt >100$~GeV, at least one of which must be a lepton. 
These events are required to have a minimum $\sumpt$ of $700$~GeV. To determine limits on the cross section for the signal production of these final states, 
this threshold is varied between $700$ and $1500$~GeV.
In making exclusion contours in the $\MD$-$\MTH$ plane, using the benchmark models described in
Section~\ref{sec:mcsamples}, a single signal region is used, defined by a $\sumpt > 1500$~GeV requirement. 

\section{Background Estimation}
\label{sec:backgrounds}

The backgrounds are estimated using a combination of data-driven and MC-based techniques. The dominant Standard Model 
sources of background are: $W+$jets,\\
$Z/\gamma^*+$jets, $t\bar{t}$ and other non-$t\bar{t}$ multi-jet processes, subsequently referred to as multi-jet events. 
In $W+$jets, 
$Z/\gamma^*+$jets and $t\bar{t}$ processes, events are produced with real leptons, and associated additional 
high-$\pt$ jets. 
In multi-jet events, reconstructed high-$\pt$ leptons are present either due to the production of a real lepton within a jet, via semileptonic
quark decays (dominantly heavy flavour decays), or due to a jet being misreconstructed from calorimeter clusters as a high-$\pt$ electron.  
These are denoted as fake leptons while those 
originating from $\tau$-leptons or heavy gauge bosons are referred to as prompt leptons.

The contribution to the muon channel signal region from multi-jets is predicted to be negligible by MC simulations, cross-checked with data using a non-isolated muon sample with the yield extrapolated to the signal region criteria.
The multi-jet contribution to the electron channel is estimated using a data-driven matrix method, described in detail in Ref.~\cite{atlas-topxsec}. 
Using the signal region definition, a multi-jet enhanced region is defined by loosening the electron identification criterion used in the event selection from ``tight'' to ``medium''.

The numbers of data events in this looser electron sample which pass ($N_\mathrm{pass}$) and fail ($N_\mathrm{fail}$) the 
final, tighter lepton selection criteria are counted. $N_\mathrm{prompt}$ and $N_\mathrm{fake}$ are defined as the numbers of events for which the 
electrons are prompt and fake, respectively. 
The following relationships hold:

\begin{equation}
N_\mathrm{pass} = \epsilon_\mathrm{prompt} N_\mathrm{prompt} + \epsilon_\mathrm{fake} N_\mathrm{fake},
\end{equation}
\begin{equation}
N_\mathrm{fail} = (1-\epsilon_\mathrm{prompt})N_\mathrm{prompt} + (1-\epsilon_\mathrm{fake}) N_\mathrm{fake}.
\end{equation}

Simultaneous solution of these two equations gives a prediction for the number of events in data in the signal region which are events with fake leptons:

\begin{equation}
N^\mathrm{pass}_\mathrm{fake}=\epsilon_\mathrm{fake}N_\mathrm{fake} = 
\frac{N_\mathrm{fail}-(1/\epsilon_\mathrm{prompt}-1)N_\mathrm{pass}}{1/\epsilon_\mathrm{fake}-1/\epsilon_\mathrm{prompt}} \> \> .
\end{equation}

The efficiency $\epsilon_\mathrm{fake}$ is determined from a multi-jet dominated data control region defined by $300 < \sumpt < 700$~GeV and $\met<15$~GeV, in which events must have at least three reconstructed 
objects passing preselection criteria, 
in the electron channel.
This region is also considered with the electron criterion loosened to ``medium''.
The efficiency for identifying 
fakes as prompt electrons is measured as the fraction of these events which also pass the tighter electron identification requirement.
The MC simulations are used to correct the efficiency for the small fraction ($<10\%$) of prompt leptons.  No dependence on lepton $\pt$, $\sumpt$ or the choice of maximum $\met$ used to define the control region is observed.

The efficiency $\epsilon_\mathrm{prompt}$ is evaluated in a second control region, again containing at least three preselected objects, but with at least two opposite-sign electrons satisfying 
$80<m_{\ell\ell}<100$~GeV, where $m_{\ell\ell}$ denotes the dilepton invariant mass. 
The efficiency for identifying prompt electrons is obtained 
through the ratio of ``medium-medium'' to ``medium-tight'' events in this high purity control region.  

\begin{table*}[htb]
 \begin{center}
   \begin{tabular}{|c|cccc|c|} \hline
   $\sumpt$ (GeV)  &  Multi-jets                          &    $W+$jets/$t\bar t$          &    $Z+$jets          &    Total SM    &  Data          \\ \hline
  $>700$   & 137 $\pm$  10     $\pm$  45   & 371  $\pm$  10 $\pm$   77    & 119   $\pm$ 4   $\pm$ 22   & 627  $\pm$ 15   $\pm$ 92   &  586     \\ \hline
  $>800$   & 75  $\pm$  7      $\pm$  25   & 210  $\pm$  6  $\pm$   42    & 74    $\pm$ 4     $\pm$ 13   & 358  $\pm$ 10   $\pm$ 51   &  348     \\ \hline
  $>900$   & 42  $\pm$  5      $\pm$  14   & 122  $\pm$  5  $\pm$   28    & 46.9  $\pm$ 2.8   $\pm$ 8.6    & 210  $\pm$ 8    $\pm$ 33   &  196     \\ \hline
  $>1000$  & 24.6  $\pm$  4.2  $\pm$  8.0  & 73   $\pm$  3  $\pm$   17    & 22.2  $\pm$ 1.8   $\pm$ 4.5    & 119  $\pm$ 5    $\pm$ 20   &  113     \\ \hline
  $>1200$  & 8.1   $\pm$  2.5  $\pm$  2.7  & 28.5 $\pm$  1.8  $\pm$ 7.6      & 9.1   $\pm$ 1.0  $\pm$ 1.9    & 45.7 $\pm$ 3.2    $\pm$ 8.3    &   41     \\ \hline
  $>1500$  & 1.3   $\pm$  1.1  $\pm$  0.4  & 6.3  $\pm$  0.8  $\pm$ 2.5      & 2.6   $\pm$ 0.5  $\pm$ 0.5    & 10.2 $\pm$ 1.4    $\pm$ 2.6    &   8      \\
   \hline
  \end{tabular}
   \caption{Background estimation summary as a function of $\sumpt$ in the electron channel, using the methods described in the main body of this Letter, compared to data. The first quoted errors are statistical, the second systematic.  All other backgrounds considered ($WW$, $ZZ$ and $WZ$) are estimated to have
negligible contributions.\label{tab:finalbkgselfit}}
   \end{center}

\end{table*}

\begin{table*}[htb]
 \begin{center}
   \begin{tabular}{|c|ccc|c|} \hline
   $\sumpt$ (GeV)  &      $W+$jets/$t\bar t$                &    $Z+$jets       &   Total SM   &  Data          \\ \hline
     $>700$         & 236    $\pm$ 7     $\pm$ 43  & 49  $\pm$ 3  $\pm$ 11  &  285   $\pm$ 8  $\pm$ 44    &  241 \\ \hline
    $>800$          & 129  $\pm$ 4     $\pm$ 25  & 32.0  $\pm$ 2.4  $\pm$ 7.5   &  161 $\pm$ 5    $\pm$ 26    &  145 \\ \hline
    $>900$          & 71     $\pm$ 3     $\pm$ 16  & 19.5  $\pm$ 1.7  $\pm$ 5.0   &  91    $\pm$ 3  $\pm$ 16    &   78 \\ \hline
    $>1000$         & 38.9   $\pm$ 2.3   $\pm$ 8.3   & 13.1   $\pm$ 1.3  $\pm$ 3.1   &  52.0  $\pm$ 2.6  $\pm$ 8.9     &   46 \\ \hline
    $>1200$         & 9.9    $\pm$ 1.2   $\pm$ 3.6   & 4.0    $\pm$ 0.6  $\pm$ 1.2   &  14.0  $\pm$ 1.3  $\pm$ 3.8     &   15 \\ \hline
    $>1500$         & 2.2    $\pm$ 0.5   $\pm$ 1.1   & 0.6    $\pm$ 0.2  $\pm$ 0.4   &  2.8   $\pm$ 0.5  $\pm$ 1.1     &   2  \\

   \hline
  \end{tabular}
\caption{Background estimation summary as a function of $\sumpt$ in the muon channel, using the methods described in the main body of this Letter, compared to data. 
The first quoted errors are statistical, the second systematic.  All other backgrounds considered ($WW$, $ZZ$, $WZ$ and multi-jet processes) are
estimated to have negligible contributions.\label{tab:finalbkgsmufit}}
\end{center}

\end{table*}

The numbers of $Z/\gamma^*+$jets events in the signal region for each channel are estimated by measuring the ratio of the number of events in data to 
that in MC simulation in a control region with: two opposite-sign leptons (two electrons or two muons) with $80<m_{\ell\ell}<100$~GeV, at
least three preselected objects and $300<\sumpt < 700$~GeV. 
This ratio is a scaling factor 
that is then used to rescale the pure MC prediction (normalised to the next-to-next-to-leading order (NNLO) cross section) in the signal region. 
The factors derived agree with unity to within the experimental uncertainty. 

The numbers of $W+$jets and $t\bar{t}$ events in the signal region is estimated in a similar fashion, by defining a control region containing events with:
exactly one electron (or muon, separately), with $40<m_\mathrm{T}<100$~GeV, where $m_\mathrm{T}$ is the transverse mass, calculated from the 
lepton transverse momentum vector, $\vec{p}_\mathrm{T}^{\,\ell}$, and the missing transverse momentum 
vector, 
$\vec{p}_{\mathrm{T}}^{\,\mathrm{miss}}$:

\begin{equation}
m_\mathrm{T} = \sqrt{2 \cdot p_\mathrm{T}^{\,\ell} \cdot \met \cdot ( 1 - \cos(\Delta \phi ( \vec{p}_\mathrm{T}^{\,\ell} , \vec{p}_{\mathrm{T}}^{\,\mathrm {miss}} ) ) )} \> \> ,
\end{equation}

\noindent with $30 < \met <60$~GeV, at least three preselected objects 
and $300 < \sumpt < 700$~GeV.
Due to their similar behaviour in $\sumpt$, $W+$jets and $t\bar{t}$ events are treated as a single background; a 
scaling factor is derived and used to rescale the pure MC prediction (normalised to the NNLO cross section) in the signal region.
The factors derived are consistent with unity to within the experimental uncertainty. 

\section{Systematic Uncertainties}

In this analysis, the dominant sources of systematic uncertainty on the estimated background event rates are: 
choice of the control regions used to derive the background estimates (for the multi-jet and $Z+$jets backgrounds),
MC modelling uncertainties assessed using alternative samples produced with different generators (for the $Z+$jets, $W+$jets 
and $t\bar{t}$ backgrounds)
and the jet energy scale (JES). 
Other uncertainties include those on the jet energy resolution (JER), lepton reconstruction and identification, 
PDF uncertainties, the finite size of event samples in the control regions and the uncertainties in the effects of initial and final-state radiation.
For the $Z+$jets, $W+$jets and $t\bar{t}$ backgrounds the use of a control region in data to renormalise the MC predictions, as described 
in Section~\ref{sec:backgrounds}, mitigates the effects of most of the systematic uncertainties, which act primarily to vary the overall magnitude
of the predicted backgrounds, rather than their shapes.  
For the background estimates of $Z+$jets, $W+$jets and $t\bar{t}$ processes, the dominant uncertainties are those associated with the extrapolation of the background shape to the signal region, followed by 
the jet energy scale.
The sizes of the systematic uncertainties described above vary, depending on the channel and on the $\sumpt$ range of the signal region, but 
are typically $15-20\%$, except for the highest $\sumpt$ bins in which the MC event samples are smaller leading to larger statistical fluctuations. These are summarised in Tables~\ref{tab:finalbkgselfit} and \ref{tab:finalbkgsmufit}.

The JES and JER uncertainties are applied to Monte Carlo simulated jets, and are propagated throughout the analysis to assess their effect.
The JES uncertainties applied were measured using the complete 2010 dataset and the techniques described in Ref.~\cite{Aad:2011he}. 
The JER measured with 2010 data~\cite{Aad:2011he} is applied to all Monte Carlo simulated jets, with the difference between the nominal and recalibrated values 
taken as the systematic uncertainty.
Additional contributions are added to both of these uncertainties to account for the effect of high luminosity pile-up in the 2011 run.
The effect of pile-up on other analysis-level distributions was investigated and found to be negligible, as expected from the high-$\pt$ objects populating the signal region.

\begin{figure*}[!hbp]
\begin{center}
  \subfigure[]{ 
    \includegraphics[width=0.45\textwidth]{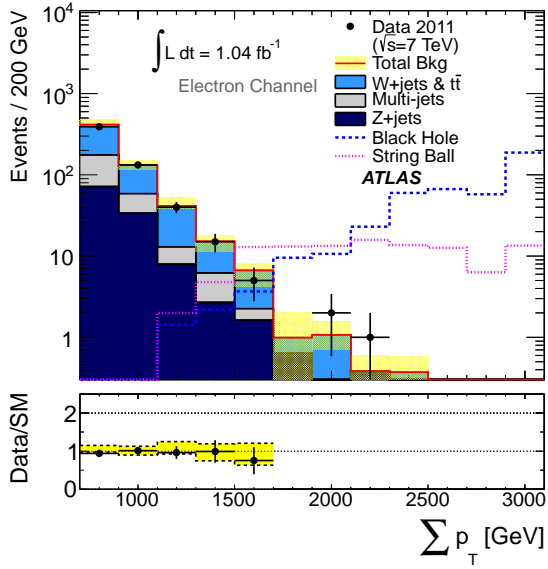}
  }
  \subfigure[]{ 
    \includegraphics[width=0.45\textwidth]{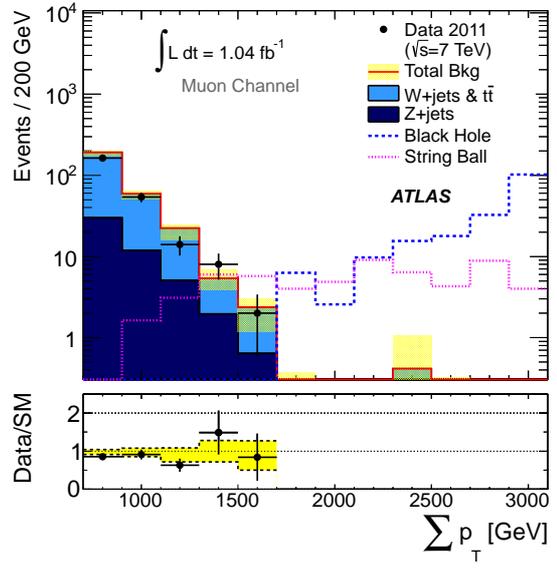}
  }
\\
  \subfigure[]{ 
    \includegraphics[width=0.45\textwidth]{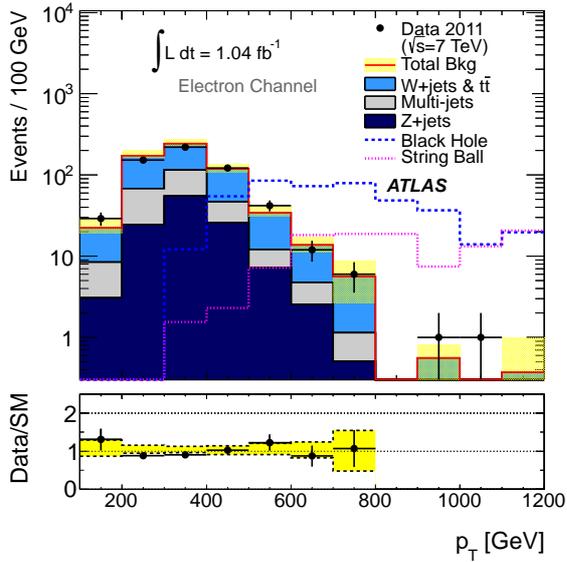}
  }
  \subfigure[]{ 
    \includegraphics[width=0.45\textwidth]{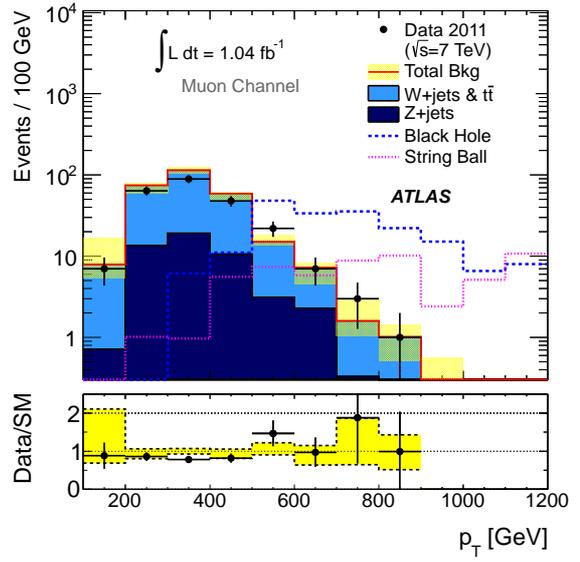}
  }
\caption{The distributions of $\sumpt$ (a), (b) and leading object $\pt$ (c), (d) for the signal region in the electron (left) and muon (right) channels. The background processes are shown according to their data-derived estimates. 
The lower panels show the ratio of the data to the expected background (points) and the uncertainty (shaded band).
The shaded band in each panel indicates the uncertainty on the expectation from the finite size of event samples, jet and lepton energy scales and resolutions.
Two representative signal
distributions are overlaid for comparison purposes.  The signal labelled ``Black Hole'' is a non-rotating black hole sample with $n=6$,
$\MD=0.8$~TeV and $\MTH=4$~TeV.  The signal labelled ``String Ball'' is a rotating string ball sample with $n=6$, $\MD=1.26$~TeV,
$\MS=1$~TeV and $\MTH=3$~TeV.  The final histogram bin 
shows the integral of all events with $\sumpt\ge  2900$~GeV (a), (b) or $\pt\ge 1100$~GeV (c), (d). }
\label{fig:finalsumpt}
\end{center}
\end{figure*}

\section{Results and Interpretation}

The observed and predicted event yields, following the estimations described in Section~\ref{sec:backgrounds}, are given in 
Tables~\ref{tab:finalbkgselfit} and~\ref{tab:finalbkgsmufit}, as a function of minimum $\sumpt$.
The distribution of 
$\sumpt$ is shown in Figure~\ref{fig:finalsumpt}, along with the distribution of the highest-$\pt$ lepton or jet.

The SM background estimates are in good agreement with the observed data, for all choices of $\sumpt$ threshold.
No excess is observed beyond the Standard Model expectation; {\it p}-values for the background-only hypothesis in the signal regions are in the range $0.43$ -- $0.47$. 
Therefore, model-independent exclusion limits are determined on the 
fiducial cross section for non-SM production of these final states, $\sigma\left(pp\to \ell X\right)$, as a function of minimum $\sumpt$.

The translation from an upper limit on the number of events to a fiducial cross section requires knowledge of the mapping (or, equivalently, 
the selection efficiency), $\epsilon_\mathrm{fid}$, from the true signal production in the fiducial region to that reconstructed. 
The true fiducial region for the electron (muon) channel is defined from simulated events with final states that pass the following requirements at generator level: 
the leading lepton is a prompt electron (muon)\footnote{Electrons (muons) originating from $\tau$-leptons, heavy gauge bosons or directly from the black hole are considered to be prompt.} within the experimental acceptance described in Section~\ref{sec:objectreconstruction}, with $\pt > 100$~GeV
and separated from jets with $\pt>20$~GeV by $\Delta R$(lepton,jet)$>0.4$; at least two additional jets or isolated leptons with $\pt > 100$~GeV are present and
$\sumpt$ is above the respective signal region threshold. 
Jets are defined using the anti-$k_t$ algorithm with $R=0.4$ on stable particles.

For the models considered, $\epsilon_\mathrm{fid}$ varies, and averages $63 \%$ for the electron channel, and $44 \%$ for the muon channel. 
The full range of $\epsilon_\mathrm{fid}$ is $57$--$67\%$ for the electron channel and $39$--$50\%$ for the muon channel.

Under the assumption of equal {\it a priori} signal model production of electrons and muons, a combined limit can also be calculated: this is a limit on the fiducial cross section for all final states with at least one lepton ($e$ or $\mu$), for which $\epsilon_\mathrm{fid}$ averages $57\%$, with a range from $50$--$61\%$.

\begin{figure}[!hbp]
\begin{center}
  \includegraphics[width=0.47\textwidth]{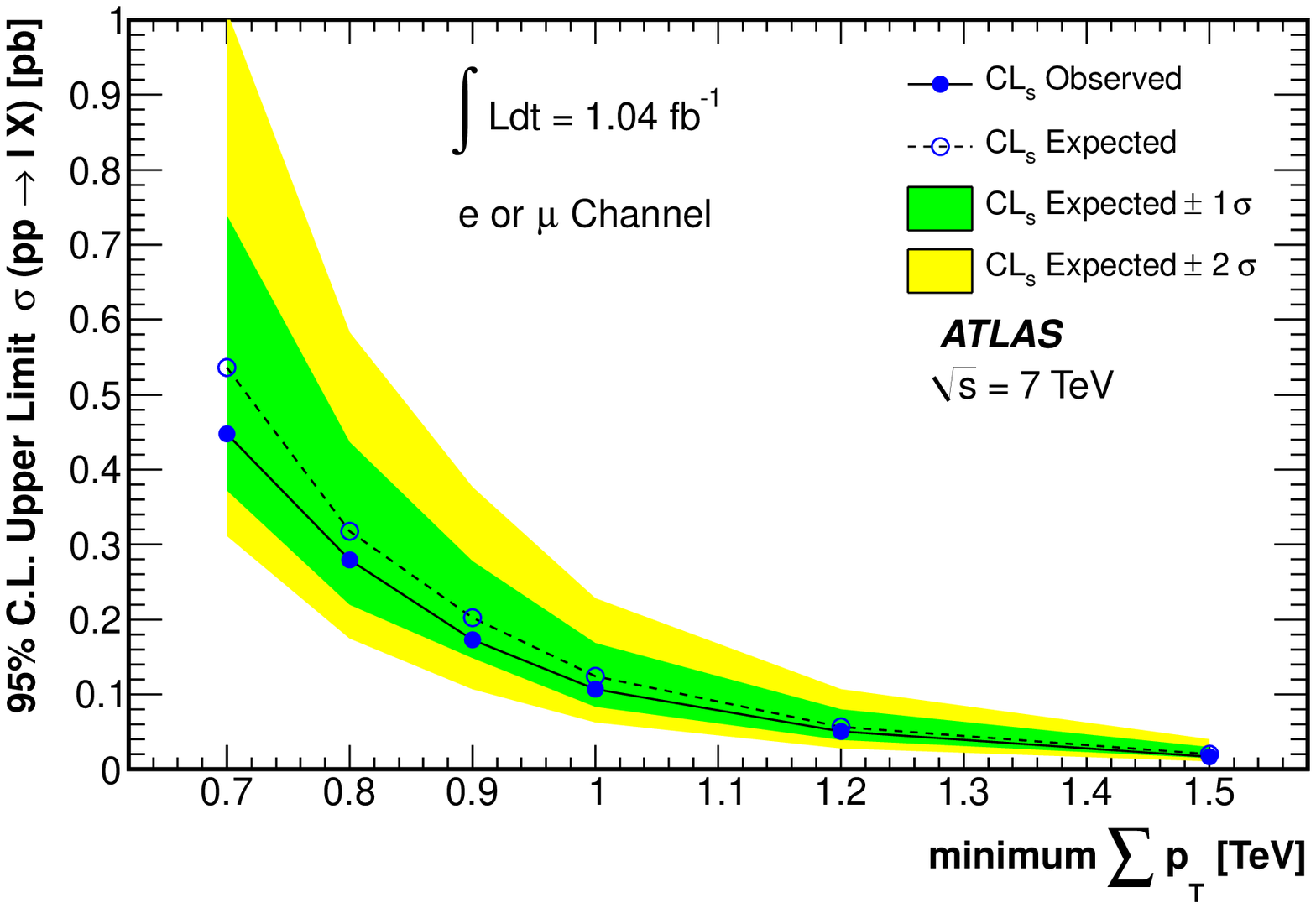}
    \caption{Upper limits on the fiducial cross sections $\sigma\left(pp\to \ell X\right)$ for the production of final states with at least three objects passing a 100~GeV $\pt$ requirement including at least one isolated lepton, and $\sumpt$ above threshold, for all final states with at least one electron or muon.
The observed and expected 95\% C.L. limits according to the CL$_{\mathrm s}$ 
prescription are shown, as well as the $1 \sigma$ and $2 \sigma$ bounds on the expected limit.}  
\label{fig:indeplimitcomb}
\end{center}
\end{figure}

For the derivation of the upper limits on the fiducial cross section, the lowest observed efficiency for each channel is used, for all signal regions. The corresponding observed and expected upper limits on the fiducial cross-section $\sigma\left(pp\to \ell X\right)$ at $95\%$ confidence level are displayed in Figure~\ref{fig:indeplimitcomb} and Table~\ref{tab:exclusion}. These exclusion regions are obtained using the CL$_{\mathrm s}$ prescription~\cite{Read:2002hq}.
For $\sumpt > 1.5$~TeV, the observed (expected) 95\% C.L. upper limit on the non-Standard Model fiducial cross section is $16.7$~fb ($20.4$~fb) for final states containing at least one electron or muon.

\begin{table}[htb]
 \begin{center}
   \begin{tabular}{|c|lll|} \hline
   $\sumpt$  & \multicolumn{3}{c|}{$\sigma\left(pp\to \ell X\right)$ $95\%$ C.L. Upper Limit (fb)}  \\ 
     & \multicolumn{3}{c|}{Observed (Expected)}  \\
  (GeV)                 & Electron  & Muon & Channels \\
		   & Channel & Channel & Combined \\ \hline
$>700$    & $282$  ($323$) &  $166$   ($233$) & $448$ ($536$)\\  
$>800$    & $179$  ($186$)   &  $117$ ($145$) & $279$ ($317$)  \\      
$>900$    & $108$  ($125$)   &  $72.6$  ($92.8$)  & $173$ ($202$)  \\    
$>1000$    & $70.9$  ($78.5$)  &  $48.2$  ($58.2$)  & $107$ ($124$)  \\
$>1200$    & $33.5$ ($38.0$)   &  $31.0$   ($28.5$) & $51.0$ ($56.8$)   \\
$>1500$    & $12.8$ ($15.4$)  &   $11.0$   ($12.3$)  & $16.7$ ($20.4$)    \\
   \hline
  \end{tabular}
   \caption{The observed and expected $95\%$ C.L. upper limits on the 
fiducial cross sections $\sigma\left(pp\to \ell X\right)$ for the production of final states with at least three objects passing a 100~GeV $\pt$ requirement including at least one isolated lepton, and $\sumpt$ above threshold, for muon and electron channels separately, and for their combination (where $l = e$ or $\mu$).
The CL$_{\mathrm s}$ method is used to obtain the limits.\label{tab:exclusion}}
   \end{center}
\end{table}

The expected and observed limits in the muon channel are slightly more stringent, due to the lower level of the SM background, in spite of the smaller efficiency and acceptance for the signal.

For the models considered, the total signal acceptance is highly model-dependent, driven primarily by the fraction of events containing a lepton in the final states, and averages about 10\% and 5\% for the (mutually exclusive) electron and muon channels respectively.
 It is lowest for the low
multiplicity, low mass states (small values of \\
$\MTH/\MD$, or $\MTH$ and $\MD$) that are theoretically or experimentally disfavoured.

\begin{figure*}[!ht]
\begin{center}
\subfigure[]{
 \includegraphics[width=0.40\textwidth]{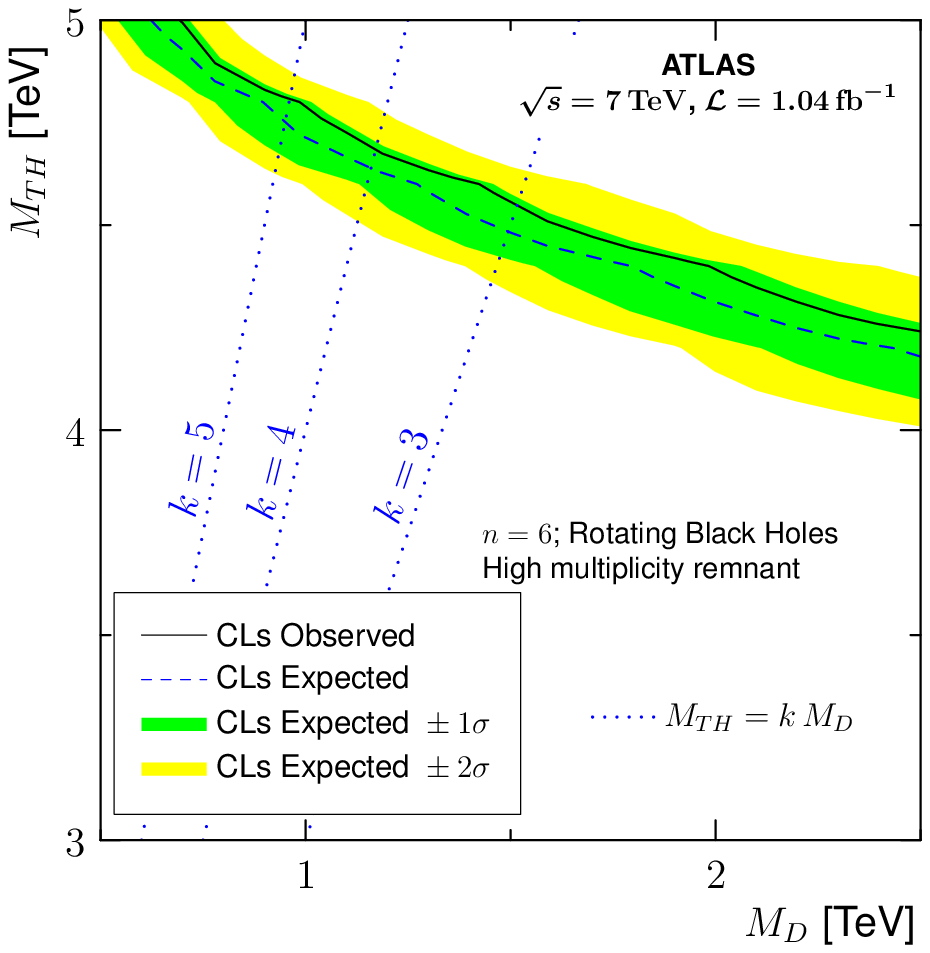}
}
\subfigure[]{
 \includegraphics[width=0.40\textwidth]{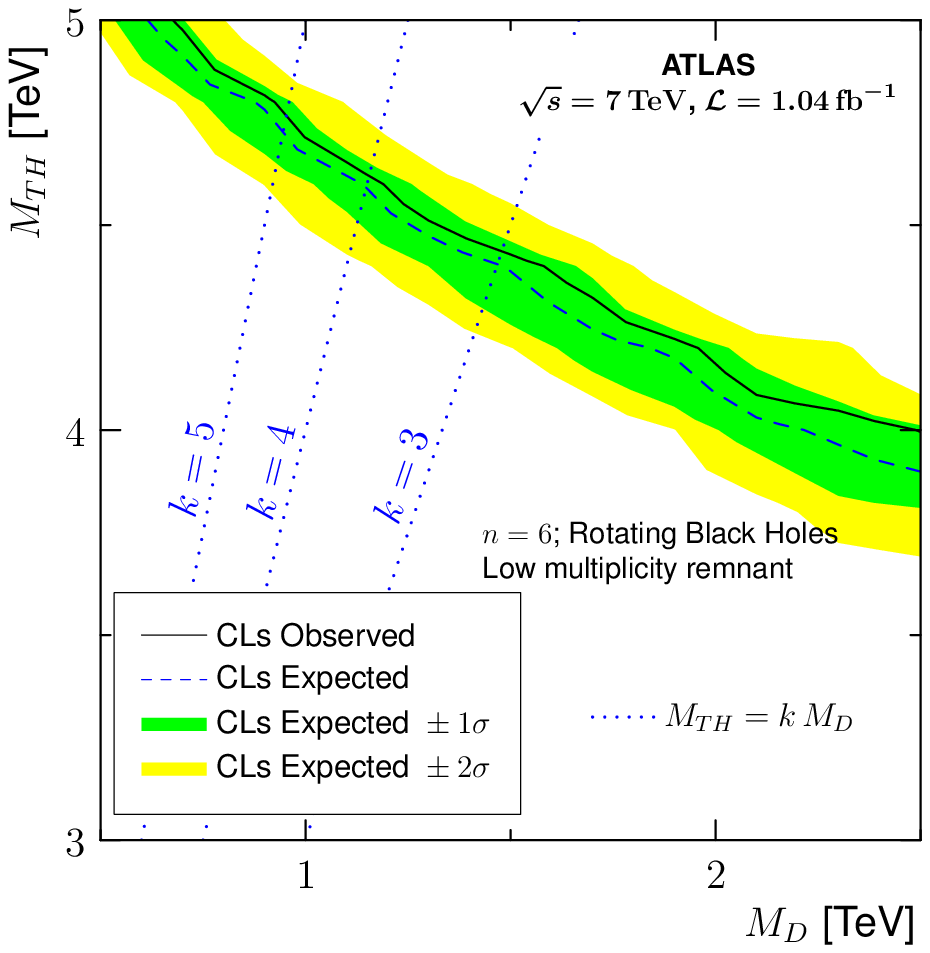}}
\caption{The exclusion limit in the $\MTH$-$\MD$ plane, with electron and muon channels combined, for rotating black hole models with six extra dimensions.  
The black hole decays result in a high-multiplicity remnant state generated with \blackmax (a), and a low-multiplicity remnant state generated by \charybdis (b).
The solid (dashed) line shows the 
observed (expected) 95\% C.L. limits, with the dark and light bands illustrating the expected $1\sigma$ and $2\sigma$ variations of
the expected limits.  The dotted lines indicate constant $k=\MTH/\MD$.
}  
\label{fig:modelBH}
 \end{center}
\end{figure*}

\begin{figure*}[!ht]
\begin{center}
\subfigure[]{
  \includegraphics[width=0.40\textwidth]{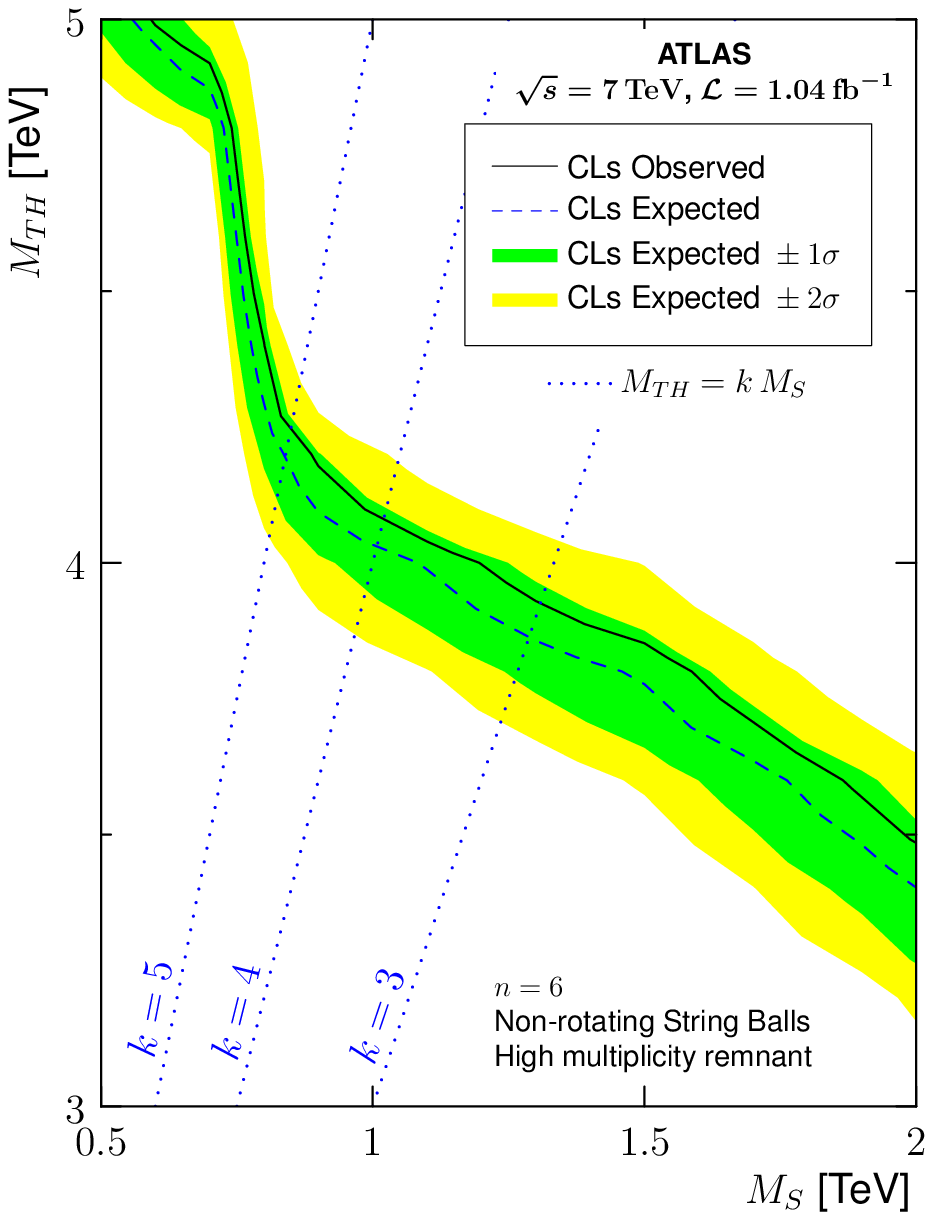}}
\subfigure[]{
  \includegraphics[width=0.40\textwidth]{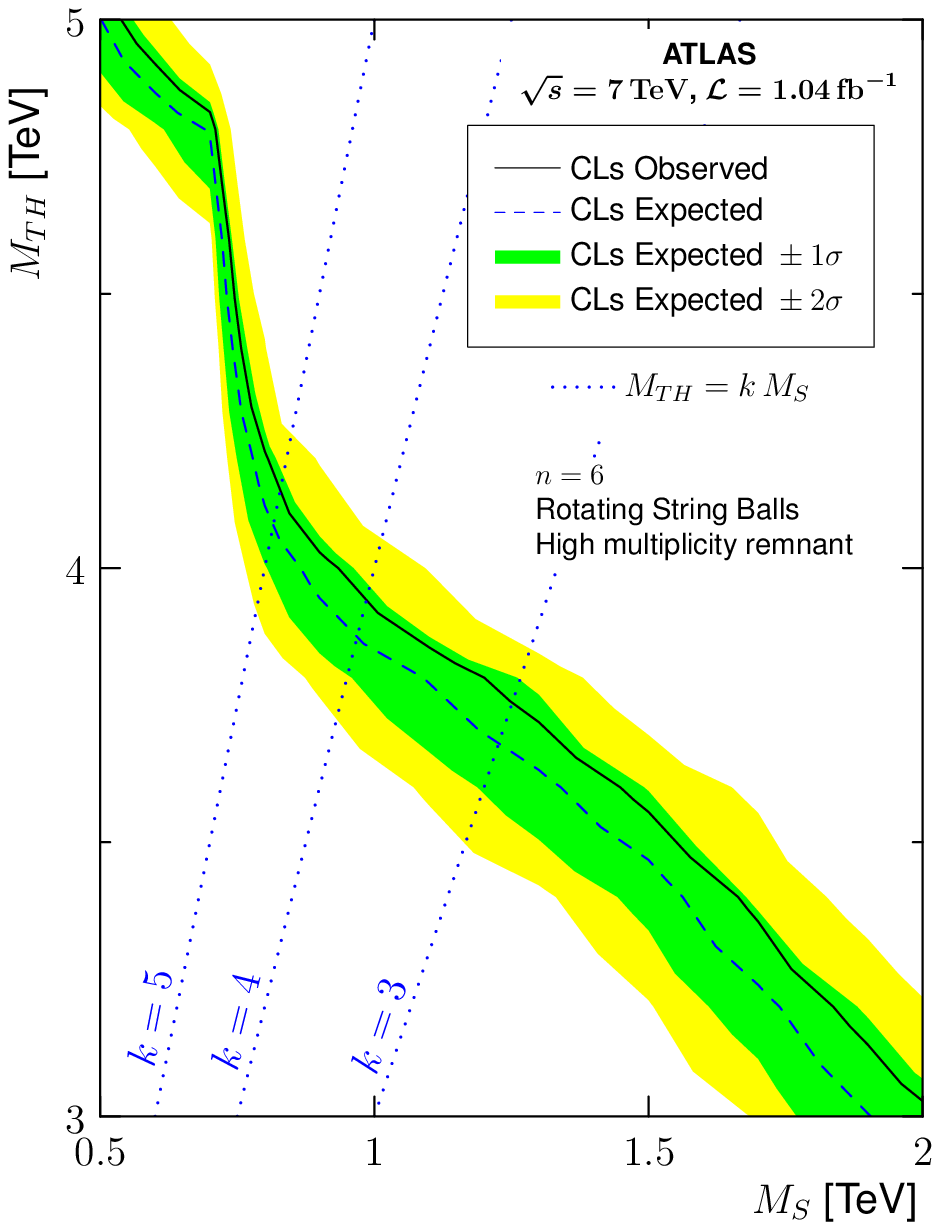}}\newline
\caption{The exclusion limit in the $\MTH$-$\MS$ plane, with electron and muon channels combined, for non-rotating (a) and rotating (b) string balls with six extra dimensions.  
The solid (dashed) line shows the 
observed (expected) 95\% C.L. limits, with the dark and light bands the expected $1$$\sigma$ and $2$$\sigma$ variations of
the expected limits.  The dotted lines indicate constant $k=\MTH/\MS$.  All samples were produced with
the \charybdis{} generator, using a high multiplicity remnant state.
} 
\label{fig:modelSB}
 \end{center}
\end{figure*}

The observed number of data events in the signal region (for $\sumpt > 1500$~GeV) along with the
background expectations are used to obtain exclusion contours in the plane of
$\MD$ and $\MTH$ for several benchmark-model gravitational states. 
No theoretical uncertainty on signal
prediction is assessed; that is, the exclusion limits are set for the exact
benchmark models as implemented in the \blackmax{} and \charybdis{} generators.
In deriving the exclusion contours, the uncertainty in the integrated luminosity and the statistical and experimental systematic uncertainties in the signal acceptances are included, and are found to be less than 10\% in total.
Some of the theoretical uncertainties, such as the effects of black hole rotation, or spin, are discussed in 
Section~\ref{sec:introduction}.  One of the more significant theoretical uncertainties is that associated with the
decay of the state as its mass approaches $\MD$.  
A common prescription is 
to end thermal emissions at a mass close to $\MD$, at which
point the state decays immediately to a remnant state, the multiplicity of which is uncertain. 
The efficiency of the event selection for searches for strong gravitational states could differ 
significantly according to the remnant model choice, particularly for samples in which a limited number of Hawking emissions are anticipated, 
motivating the consideration of multiple remnant models.  The requirement of only three high-$\pt$ objects for this analysis mitigates
the dependence of the selection efficiency, and resulting cross section limits, on the modelling of the remnant decays. 

The $95\%$ exclusion contours in the $\MD$-$\MTH$ plane ($\MS$-$\MTH$ plane for string balls) for different models
are obtained using the CL$_\mathrm{s}$ prescription. Figure~\ref{fig:modelBH} shows 
exclusion contours for rotating black hole benchmark models with high- and low-multiplicity remnant decays. Their comparison allows an assessment of the effect of this modelling uncertainty on the analysis, which is inevitably greatest in the regime of low $\MTH/\MD$.
Limits for rotating and non-rotating string ball models are shown in Figure~\ref{fig:modelSB}. 
The behaviour of the contours observed at high values of $\MTH/\MS$ are due to a step decrease in the gradient of the string ball cross section, $d \sigma_\mathrm{SB} / d \MS$ 
above a value of $\MTH = \MS/g_S^{2}$. 
The string ball models illustrated were simulated using a high-multiplicity remnant model.

\section{Summary}

A search for microscopic black holes and string ball states in ATLAS using a
total integrated luminosity of {\lumi}~fb$^{-1}$ was presented.
The search has considered final states with three or more high transverse
momentum objects, at least one of which was required to be a lepton (electron or muon).
No deviation from the Standard Model was observed in either the electron or
the muon channels. Consequently, limits are set on TeV-scale gravity models,
interpreted in a two-dimensional parameter grid of benchmark models in the $\MD$-$\MTH$ plane. 
Upper limits, at 95\% C.L., are set on the fiducial cross-sections for new physics production 
of high-$\sumpt$ multi-object final states containing a high-$\pt$ ($>100$~GeV) isolated lepton within the experimental acceptance. For final states with $\sumpt>1.5$~TeV, a limit of 16.7~fb is set.

\section{Acknowledgements}

We thank CERN for the very successful operation of the LHC, as well as the
support staff from our institutions without whom ATLAS could not be
operated efficiently.

We acknowledge the support of ANPCyT, Argentina; YerPhI, Armenia; ARC,
Australia; BMWF, Austria; ANAS, Azerbaijan; SSTC, Belarus; CNPq and FAPESP,
Brazil; NSERC, NRC and CFI, Canada; CERN; CONICYT, Chile; CAS, MOST and NSFC,
China; COLCIENCIAS, Colombia; MSMT CR, MPO CR and VSC CR, Czech Republic;
DNRF, DNSRC and Lundbeck Foundation, Denmark; EPLANET and ERC, European Union;
IN2P3-CNRS, CEA-DSM/IRFU, France; GNAS, Georgia; BMBF, DFG, HGF, MPG and AvH
Foundation, Germany; GSRT, Greece; ISF, MINERVA, GIF, DIP and Benoziyo Center,
Israel; INFN, Italy; MEXT and JSPS, Japan; CNRST, Morocco; FOM and NWO,
Netherlands; RCN, Norway; MNiSW, Poland; GRICES and FCT, Portugal; MERYS
(MECTS), Romania; MES of Russia and ROSATOM, Russian Federation; JINR; MSTD,
Serbia; MSSR, Slovakia; ARRS and MVZT, Slovenia; DST/NRF, South Africa;
MICINN, Spain; SRC and Wallenberg Foundation, Sweden; SER, SNSF and Cantons of
Bern and Geneva, Switzerland; NSC, Taiwan; TAEK, Turkey; STFC, the Royal
Society and Leverhulme Trust, United Kingdom; DOE and NSF, United States of
America.

The crucial computing support from all WLCG partners is acknowledged
gratefully, in particular from CERN and the ATLAS Tier-1 facilities at
TRIUMF (Canada), NDGF (Denmark, Norway, Sweden), CC-IN2P3 (France),
KIT/GridKA (Germany), INFN-CNAF (Italy), NL-T1 (Netherlands), PIC (Spain),
ASGC (Taiwan), RAL (UK) and BNL (USA) and in the Tier-2 facilities
worldwide.

\bibliographystyle{atlasnote}
\bibliography{LJPaper}

\clearpage
\onecolumn
\begin{flushleft}
{\Large The ATLAS Collaboration}

\bigskip

G.~Aad$^{\rm 48}$,
B.~Abbott$^{\rm 110}$,
J.~Abdallah$^{\rm 11}$,
A.A.~Abdelalim$^{\rm 49}$,
A.~Abdesselam$^{\rm 117}$,
O.~Abdinov$^{\rm 10}$,
B.~Abi$^{\rm 111}$,
M.~Abolins$^{\rm 87}$,
O.S.~AbouZeid$^{\rm 157}$,
H.~Abramowicz$^{\rm 152}$
% ...
\bigskip

% ...
$^{10}$ Institute of Physics, Azerbaijan Academy of Sciences, Baku, Azerbaijan\\
$^{11}$ Institut de F\'isica d'Altes Energies and Departament de F\'isica de la Universitat Aut\`onoma  de Barcelona and ICREA, Barcelona, Spain\\
% ...
$^{48}$ Fakult\"{a}t f\"{u}r Mathematik und Physik, Albert-Ludwigs-Universit\"{a}t, Freiburg i.Br., Germany\\
$^{49}$ Section de Physique, Universit\'e de Gen\`eve, Geneva, Switzerland\\
% ...
$^{87}$ Department of Physics and Astronomy, Michigan State University, East Lansing MI, United States of America\\
% ...
$^{110}$ Homer L. Dodge Department of Physics and Astronomy, University of Oklahoma, Norman OK, United States of America\\
$^{111}$ Department of Physics, Oklahoma State University, Stillwater OK, United States of America\\
% ...
$^{117}$ Department of Physics, Oxford University, Oxford, United Kingdom\\
% ...
$^{152}$ Raymond and Beverly Sackler School of Physics and Astronomy, Tel Aviv University, Tel Aviv, Israel\\
% ...
$^{157}$ Department of Physics, University of Toronto, Toronto ON, Canada\\
% ...
\end{flushleft}

\end{document}